\documentclass{emulateapj}

\begin{document}

\title{FAR-ULTRAVIOLET EMISSION-LINE MORPHOLOGIES OF THE SUPERNOVA REMNANT G65.3+5.7}

\author{I.-J. Kim\altaffilmark{1}}
\author{K.-I. Seon\altaffilmark{1}}
\author{K.-W. Min\altaffilmark{2}}
\author{J.-H. Shinn\altaffilmark{1}}
\author{W. Han\altaffilmark{1}}
\author{J. Edelstein\altaffilmark{3}}

\altaffiltext{1}{Korea Astronomy and Space Science Institute, 305-348, Daejeon, South Korea; ijkim@kasi.re.kr}
\altaffiltext{2}{Korea Advanced Institute of Science and Technology, 305-701, Daejeon, South Korea}
\altaffiltext{3}{Space Sciences Laboratory, University of California, Berkeley, CA 94720}

\begin{abstract}
We present the first far-ultraviolet (FUV) emission-line morphologies of the whole region of the supernova remnant (SNR) \object{G65.3+5.7} using the FIMS/SPEAR data. The morphologies of the \ion{C}{4} $\lambda\lambda$1548, 1551, \ion{He}{2} $\lambda$1640, and \ion{O}{3}] $\lambda\lambda$1661, 1666 lines appear to be closely related to the optical and/or soft X-ray images obtained in previous studies. Dramatic differences between the \ion{C}{4} morphology and the optical [\ion{O}{3}] $\lambda$5007 image provide clues to a large resonant-scattering region and a foreground dust cloud. The FUV morphologies also reveal the overall distribution of various shocks in different evolutionary phases and an evolutionary asymmetry between the east and the southwest sides in terms of Galactic coordinates, possibly due to a Galactic density gradient in the global scale. The relative X-ray luminosity of \object{G65.3+5.7} to \ion{C}{4} luminosity is considerably lower than those of the \object{Cygnus Loop} and the \object{Vela} SNRs. This implies that \object{G65.3+5.7} has almost evolved into the radiative stage in the global sense and supports the previous proposal that \object{G65.3+5.7} has lost its bright X-ray shell and become a member of mixed-morphology SNRs as it has evolved beyond the adiabatic stage.
\end{abstract}


\section{INTRODUCTION}

The supernova remnant (SNR) \object{G65.3+5.7} was first identified in an optical emission-line survey \citep{gul77} although spectroscopic work on portions of the remnant had been previously performed \citep{sab76}. Especially in the [\ion{O}{3}] $\lambda$5007 emission line, it appeared to have prominent filamentary features with an angular size of 4$\arcdeg$ $\times$ 3.3$\arcdeg$. By radio continuum observations, \citet{rei79} found a shell-type radio morphology coinciding with the bright optical filaments and estimated the distance to this target to be $\sim$900 pc. Recently, \citet{mav02} presented deep optical images of [\ion{O}{2}] $\lambda$3727 and [\ion{O}{3}] $\lambda$5007, and deep long-slit spectra at several different positions. Based on significant morphological differences between the two emission-line images and high [\ion{O}{3}]/H$\beta$ ratios (6.4--110) in most of the regions, they suggested prevalence of incomplete recombination zones behind its shocks. In supplementary work, \citet{bou04} estimated a global expansion velocity of 124--187 km s$^{-1}$ from the [\ion{O}{3}] line profiles. \citet{xia09} obtained high spatial resolution images of the radio continuum at two frequencies (4.8 and 2.6 GHz) and performed detailed analyzes of the radio properties. They also investigated \ion{H}{1} 21 cm data of the \object{G65.3+5.7} region. They concluded that \object{G65.3+5.7} has almost entered the cooling phase as typical of evolved shell-type SNRs and may be expanding in a pre-blown cavity. \object{G65.3+5.7} is also a bright soft X-ray source. Using ROSAT PSPC data, \citet{she04} found centrally bright X-ray morphologies and concluded that \object{G65.3+5.7} is a thermal composite (also known as mixed-morphology) SNR in the radiative phase. By the X-ray spectral fitting, they estimated the electron temperature to be 2.5--3.5 $\times$ 10$^{6}$ K and concluded that the foreground neutral hydrogen column density ranged from 8 $\times$ 10$^{19}$ to 1.3 $\times$ 10$^{20}$ cm$^{-2}$.

Judging from the fact that \object{G65.3+5.7} has little foreground extinction and that it is evolving into the cooling phase, as revealed in previous studies, \object{G65.3+5.7} is expected to be observed brightly in the far-ultraviolet (FUV; 900--1800 \AA) wavelength domain. Especially, in incomplete radiative shocks, the FUV lines are stronger than the low-ionization optical lines, such as [\ion{S}{2}] and Hydrogen Balmer lines \citep{ray88}. Thus, FUV observations are particularly important in understanding \object{G65.3+5.7}, which has incomplete recombination zones in most of its shocks. In this paper, we present the first FUV emission-line morphologies of the whole region of \object{G65.3+5.7} using data from the Far-Ultraviolet IMaging Spectrograph (FIMS), also known as Spectroscopy of Plasma Evolution from Astrophysical Radiation (SPEAR). The morphologies of the \ion{C}{4} $\lambda\lambda$1548, 1551, \ion{He}{2} $\lambda$1640, and \ion{O}{3}] $\lambda\lambda$1661, 1666 lines show close relations with the optical and/or X-ray images, and the overall distribution of various shocks in different evolutionary stages. A quantitative comparison of total \ion{C}{4} and X-ray luminosities also shows that \object{G65.3+5.7} has almost evolved into the radiative stage in the global sense.

\section{OBSERVATIONS AND DATA REDUCTION}

FIMS/SPEAR is the primary payload of the first Korean Science and Technology Satellite, {\it STSAT-1}, a micro-satellite launched on 2003 September 27. FIMS/SPEAR was designed to observe large-scale diffuse FUV emission lines from the interstellar medium (ISM). FIMS/SPEAR consists of dual FUV imaging spectrographs: the short wavelength channel (S-channel; 900--1150 \AA, 4.0$\arcdeg$ $\times$ 4.6$\arcmin$ field of view) and the long wavelength channel (L-channel; 1340--1750 \AA, 7.4$\arcdeg$ $\times$ 4.3$\arcmin$ field of view), with $\lambda / \Delta \lambda \sim 550$ spectral resolution and 5$\arcmin$ angular resolution. The instrument, its on-orbit performance, and the basic processing of the data are described in detail in Edelstein et al. (2006a, 2006b).

We used the L-channel data from a total of 83 orbits obtained in the sky survey observational mode. Of the 83 total orbits, 76 orbits were observed in the 10\% shutter transmission mode, and 7 orbits were observed in the 100\% mode. The exposures taken with the 10\% mode were scaled by the factor of 0.1, as in \citet{seo10}. The data includes a total of 7260 events for the area including \object{G65.3+5.7}. The limited attitude accuracy of the satellite was augmented by automated software correction using the positions of the bright stars listed in the {\it TD-1} catalog \citep{tho78}. The positions of the reference stars were corrected to be accurate within 5$\arcmin$ in the present study. Then, the position errors of all photons were corrected by linearly interpolating from those of the bright stars. This method is similar to that used in \citet{seon06}. To obtain images and spectra, we have adopted the HEALPix scheme \citep{gor05} with the resolution parameter \texttt{Nside} = 1024, corresponding to a pixel size of $\sim$3.4$\arcmin$, and 2 \AA{} wavelength bins. The effects of bright stars were further reduced by masking the pixels of the elliptical region around the stars, based on the scattering pattern of the slit image. We identified a total of 17 stars in the area of present concern using the {\it TD-1} catalog. The masked positions are indicated in Figure 1.

\begin{figure}
\epsscale{0.55} \plotone{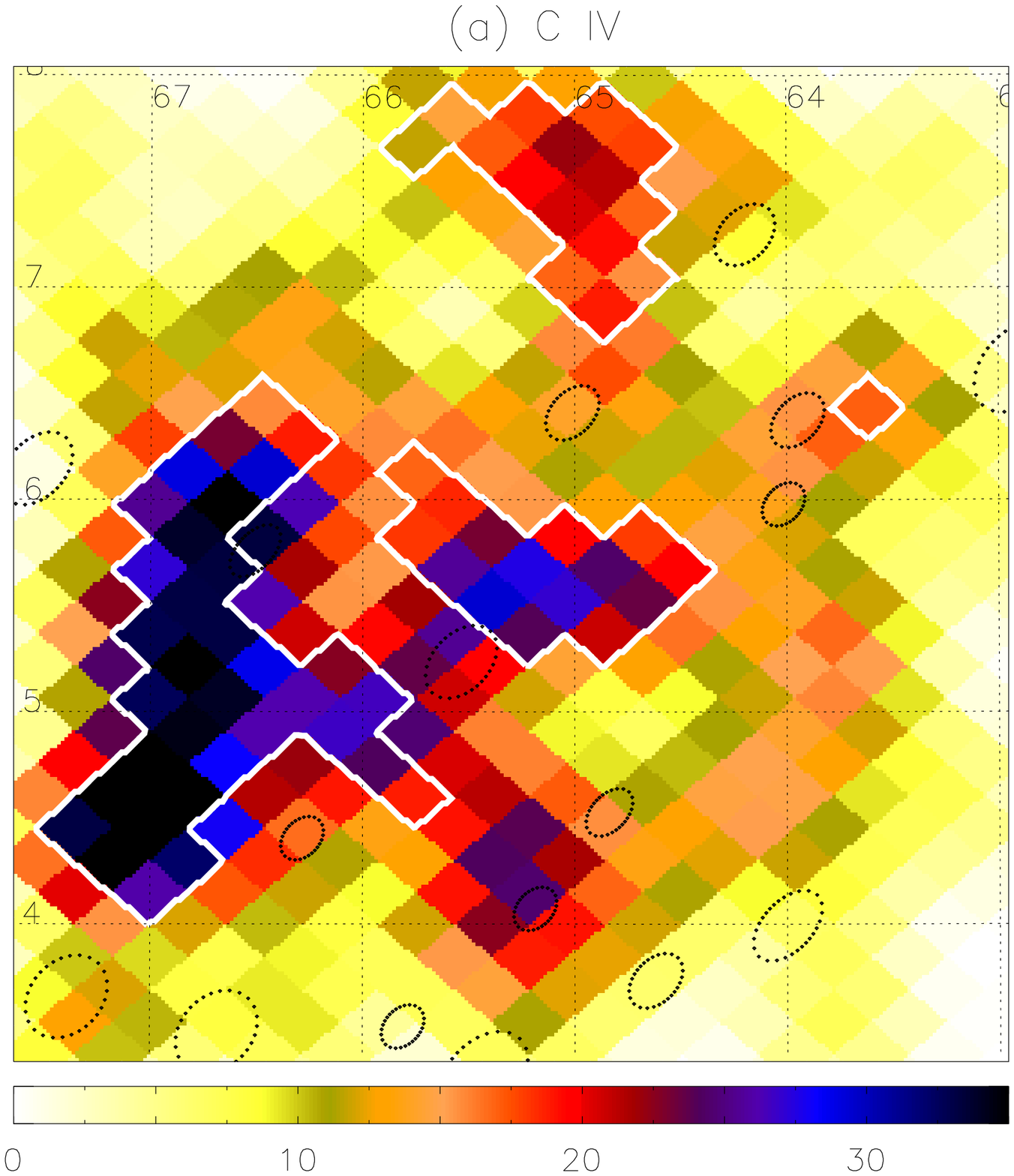}\quad\plotone{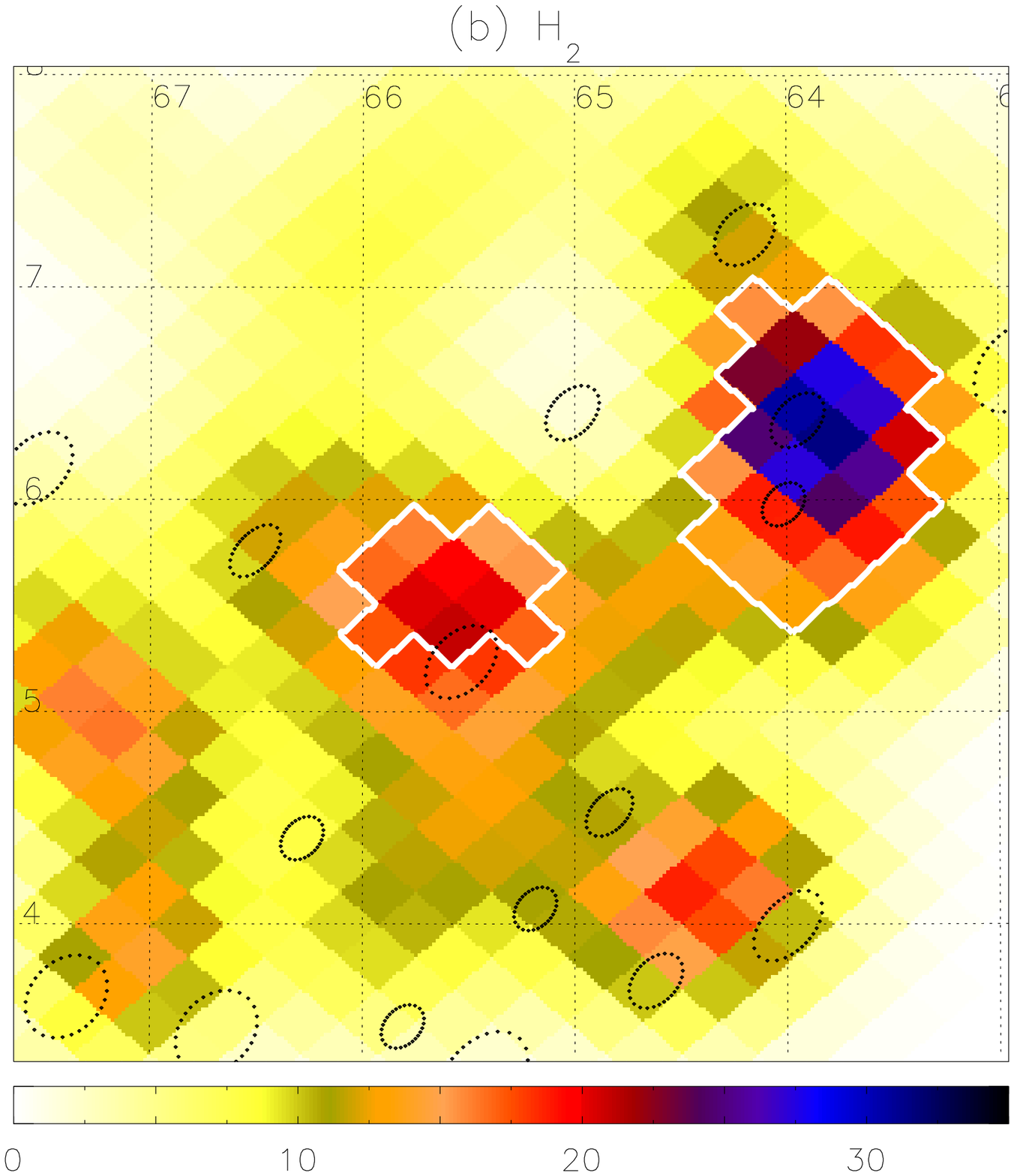}\plotone{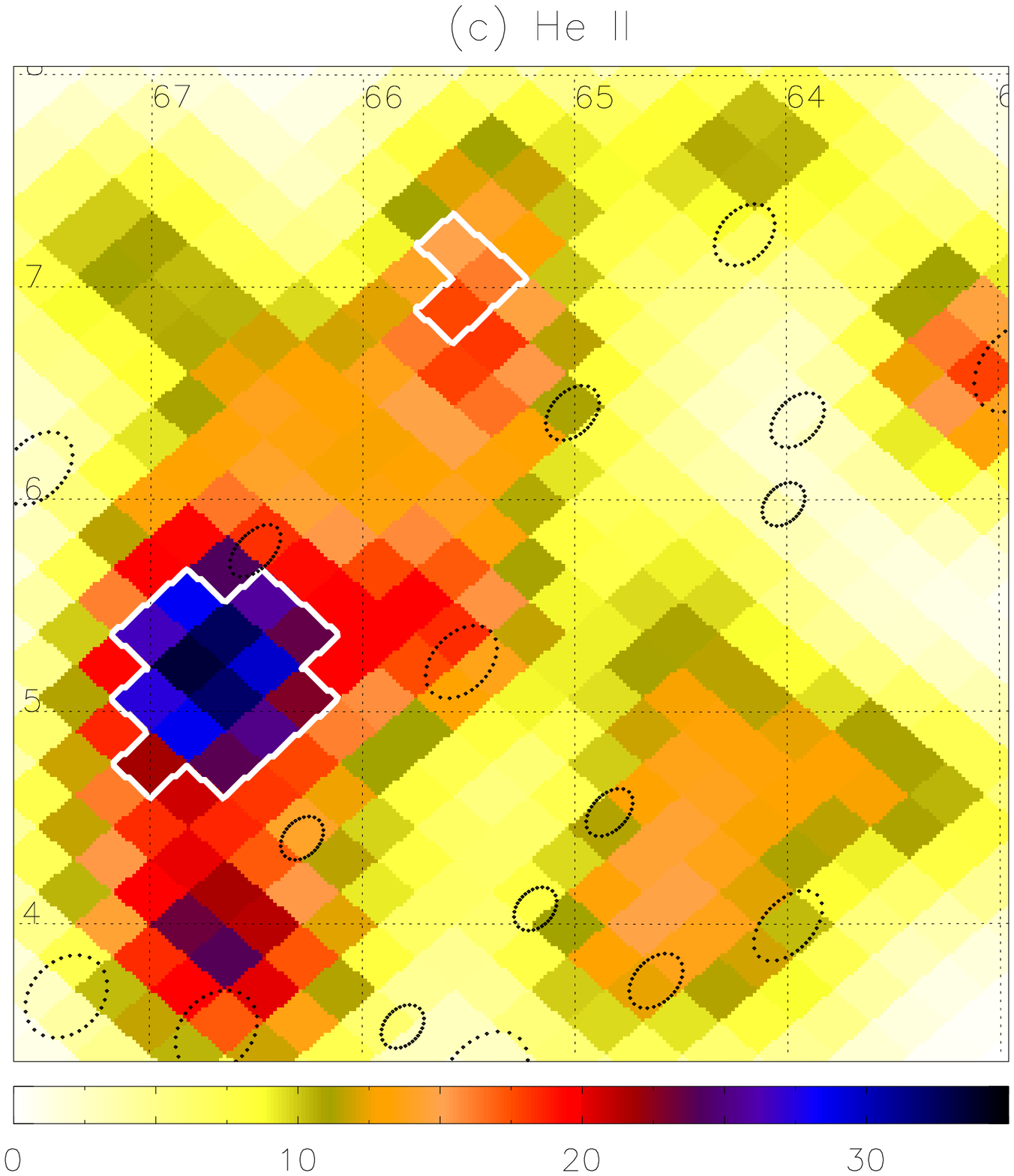}\quad\plotone{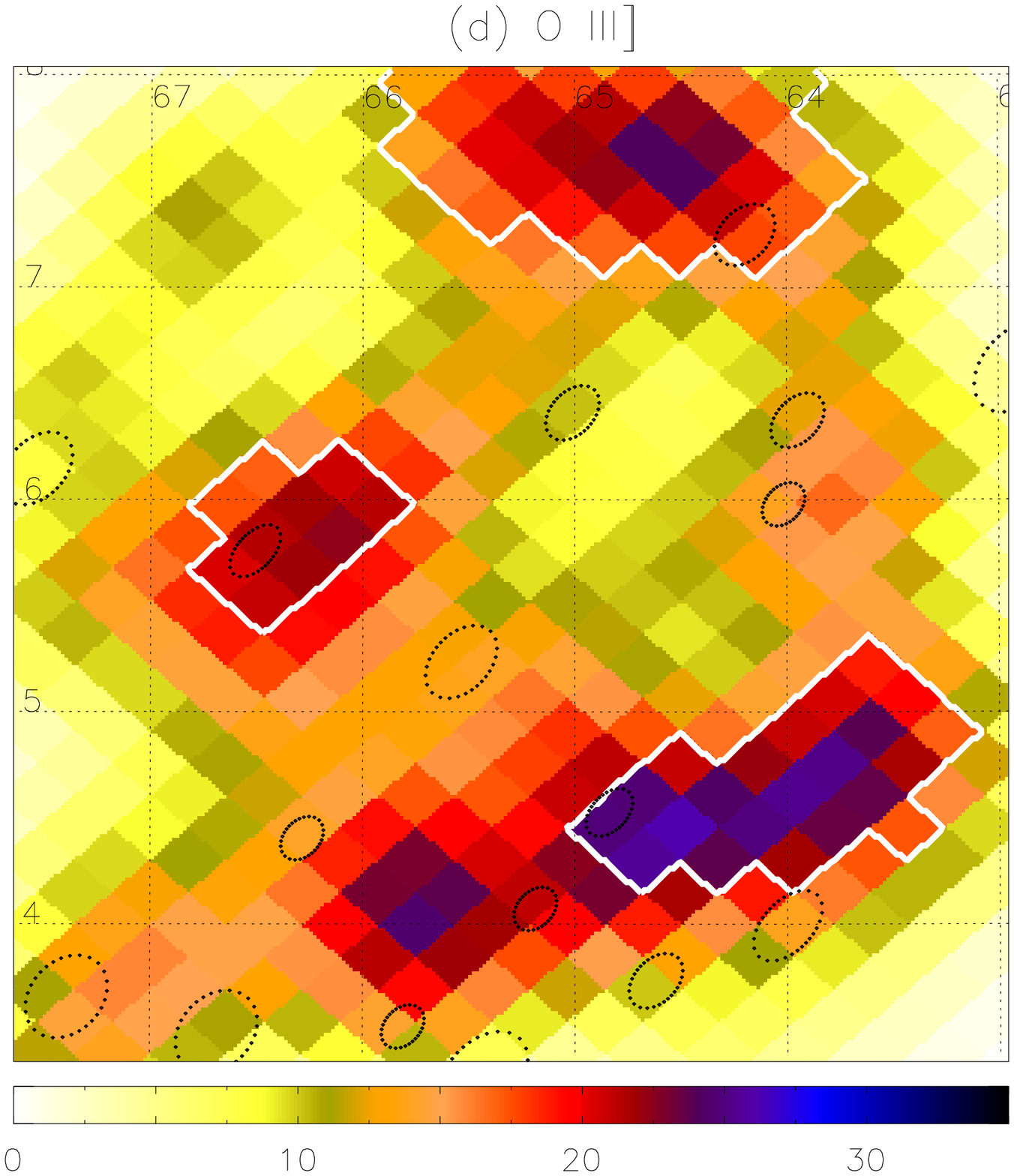} \caption{FIMS/SPEAR (a) \ion{C}{4} $\lambda\lambda$1548, 1551, (b) H$_{2}$ $\lambda$1608, (c) \ion{He}{2} $\lambda$1640, and (d) \ion{O}{3}] $\lambda\lambda$1661, 1666 emission-line images of SNR G65.3+5.7 in the Galactic coordinates. The units of the color bars are $10^{-7}$ ergs s$^{-1}$ cm$^{-2}$ sr$^{-1}$. The values are not corrected for interstellar extinction. The dotted-line ellipses are the regions masked to remove stellar contamination. The solid lines enclose the regions where signal-to-noise ratios are $>$1.5. \label{fig1}}
\end{figure}

\section{DATA ANALYSIS AND RESULTS}

To make emission-line images, four portions (1530--1570, 1592--1624, 1624--1656, and 1650--1680 \AA, corresponding to the spectral regions around the emission lines of \ion{C}{4}, H$_{2}$, \ion{He}{2}, and \ion{O}{3}], respectively) were taken from the L-channel spectrum and each portion was fitted with a constant continuum plus convolved model lines of Gaussian shape for each pixel. The centers and widths of the Gaussian functions were fixed using the calibrated line centers and the FIMS/SPEAR spectral resolutions, respectively. The \ion{C}{4} $\lambda\lambda$1548, 1551 doublet lines have intrinsic line ratio of 2:1, but resonant scattering makes the ratio approach to 1:1 in an optically thick case \citep{lon92}. We assumed a moderate case of the 1.5:1 line ratio. Actually, the results are not largely different from those of both extreme cases because the spectral resolution of FIMS/SPEAR is unable to resolve the \ion{C}{4} doublet lines. The \ion{O}{3}] lines were assumed to be formed in a doublet at 1661 and 1666 \AA{} in the 0.41:1 ratio of their statistical weights. Before fitting, we increased the pixel size by 4 times (\texttt{Nside} = 256, corresponding to a pixel size of $\sim$13.7$\arcmin$) for better statistics. The resulting images were sparse and some pixels contain only a few event-counts because of the limited exposure time. To obtain significant morphologies, the images were then smoothed using the spherical-harmonics transformation of a Fisher-von Mises function, the mathematical version of a Gaussian function in spherical space \citep{seo06}. The full widths at a half-maximum of the Gaussian kernel are 35.0$\arcmin$ for \ion{C}{4} and 50.0$\arcmin$ for the other lines, which are appropriate smoothing lengths for preserving the emission pattern in each original unsmoothed image.

Figure 1 shows the final results of the four emission-line images in the Galactic coordinates. The signal-to-noise ratio (S/N) of each pixel is quite low because of the limited exposure time. In the figure, the solid lines indicate the boundaries of S/N $>$1.5. However, we believe that most bright regions are real, based on close relations with the optical and/or soft X-ray images obtained in previous studies and the fact that the average spectra have higher S/N values for the corresponding regions, as will be shown later. The \ion{C}{4} image shows two bright filamentary features in the east and central regions, while the \ion{O}{3}] image shows two filamentary features in the southwest and north regions. The emission patterns for the bright (S/N $>$1.5) H$_{2}$ $\lambda$1608 and \ion{He}{2} $\lambda$1640 regions appear to be dominated by only a few pixels in the original unsmoothed images though their existence is real in the spectra. It should be noted that the bright \ion{He}{2} region corresponds well to the enhanced \ion{C}{4} region.

\begin{figure}
\epsscale{0.55} \plotone{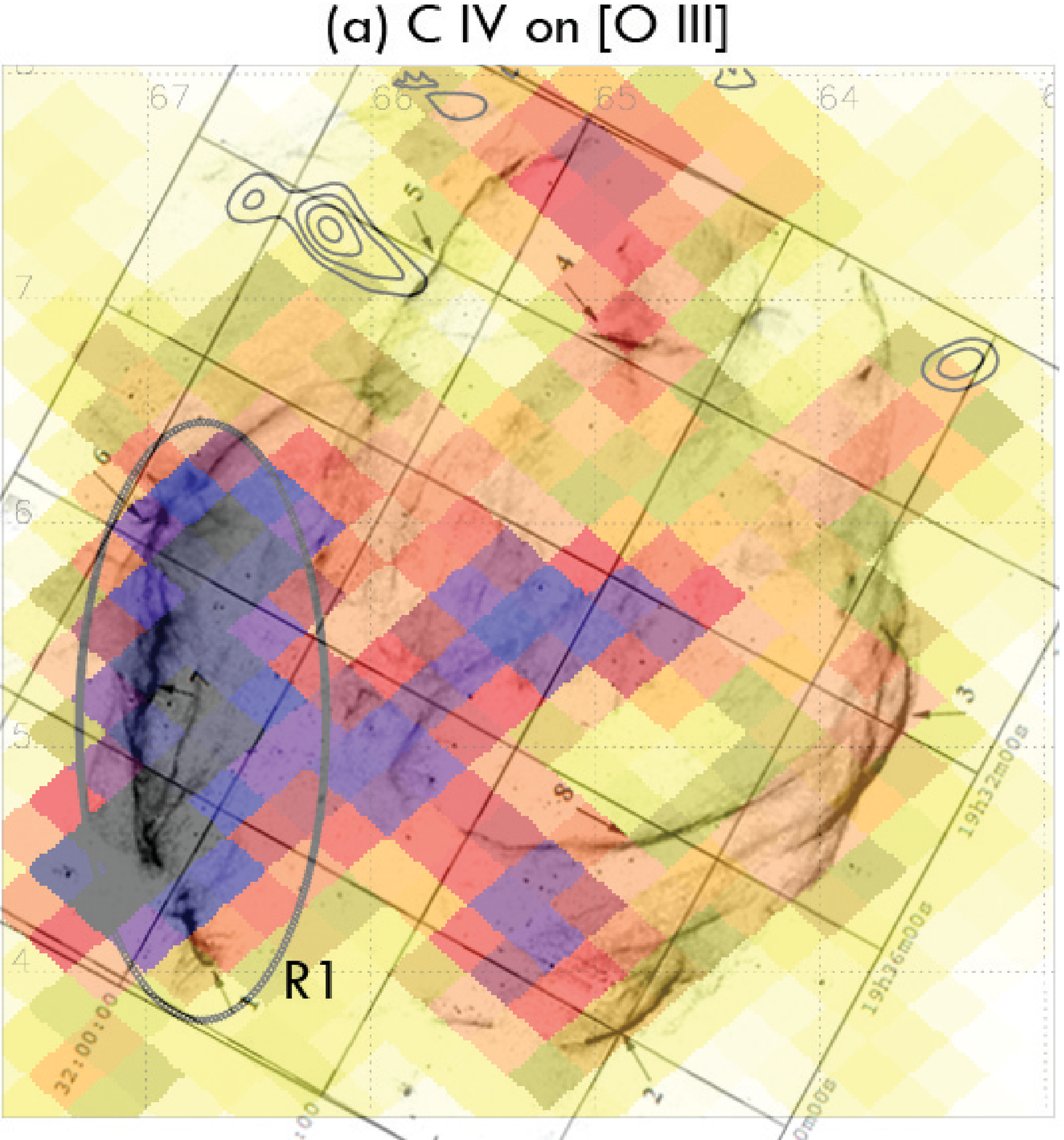}\quad\plotone{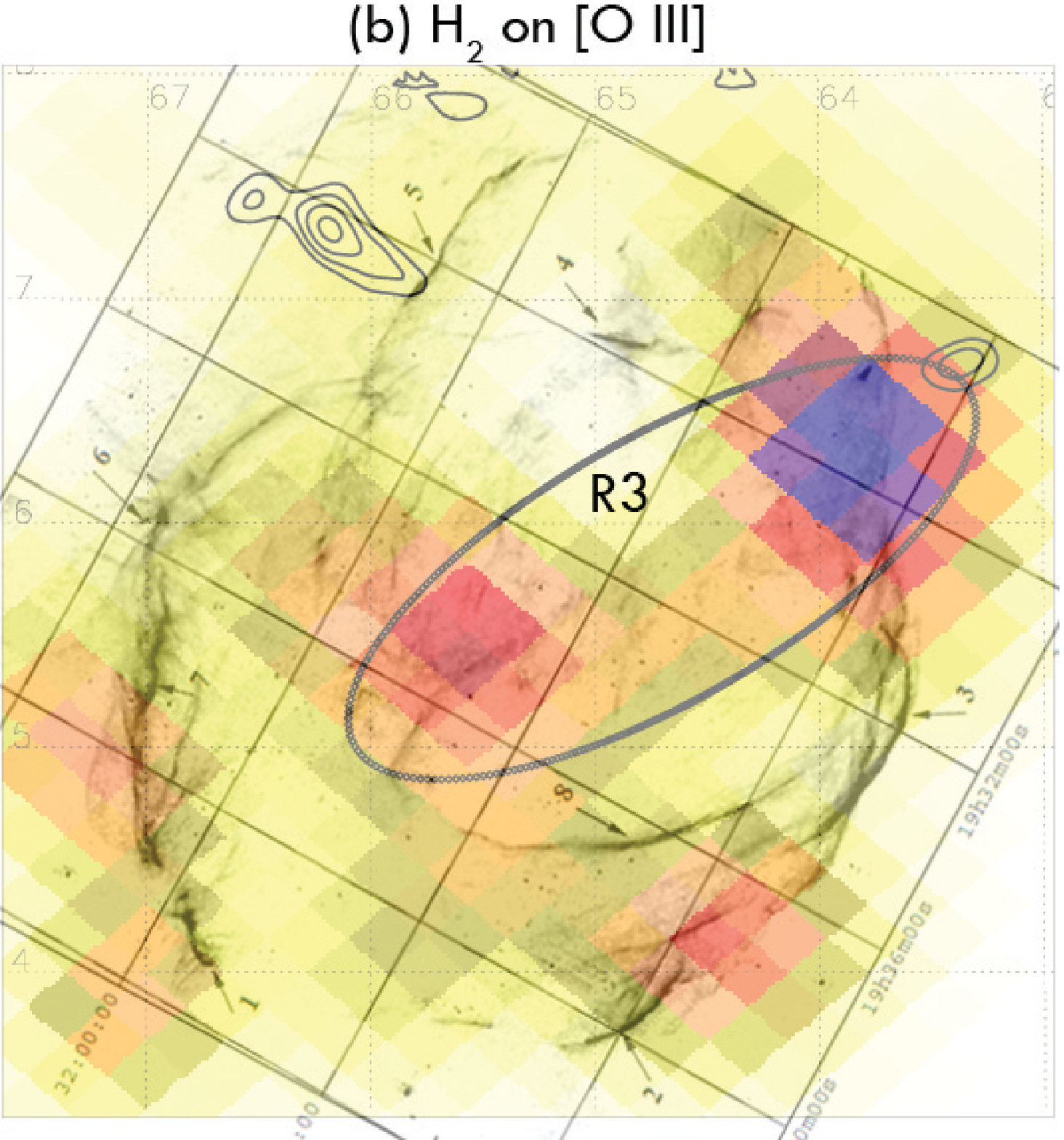}\plotone{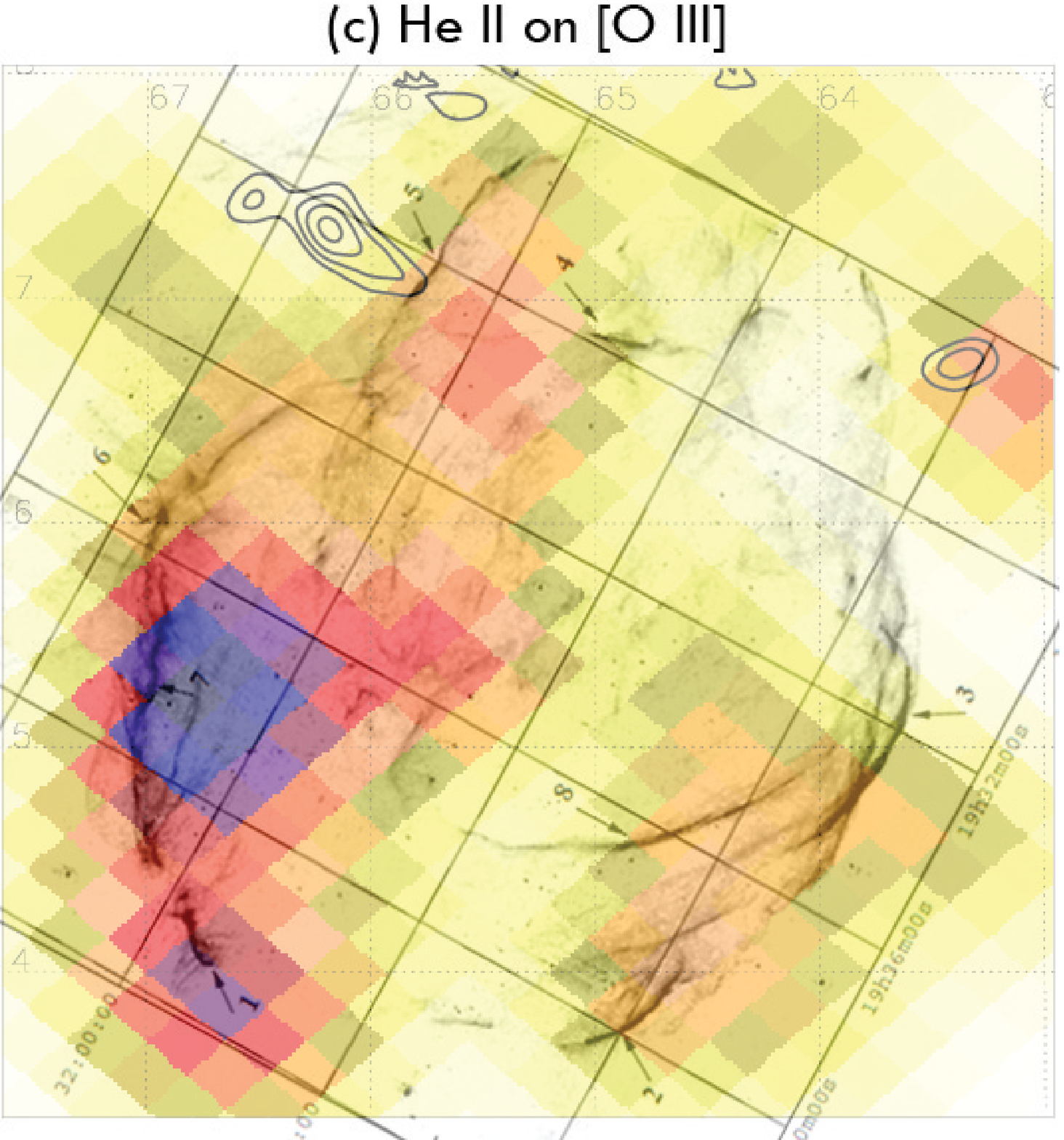}\quad\plotone{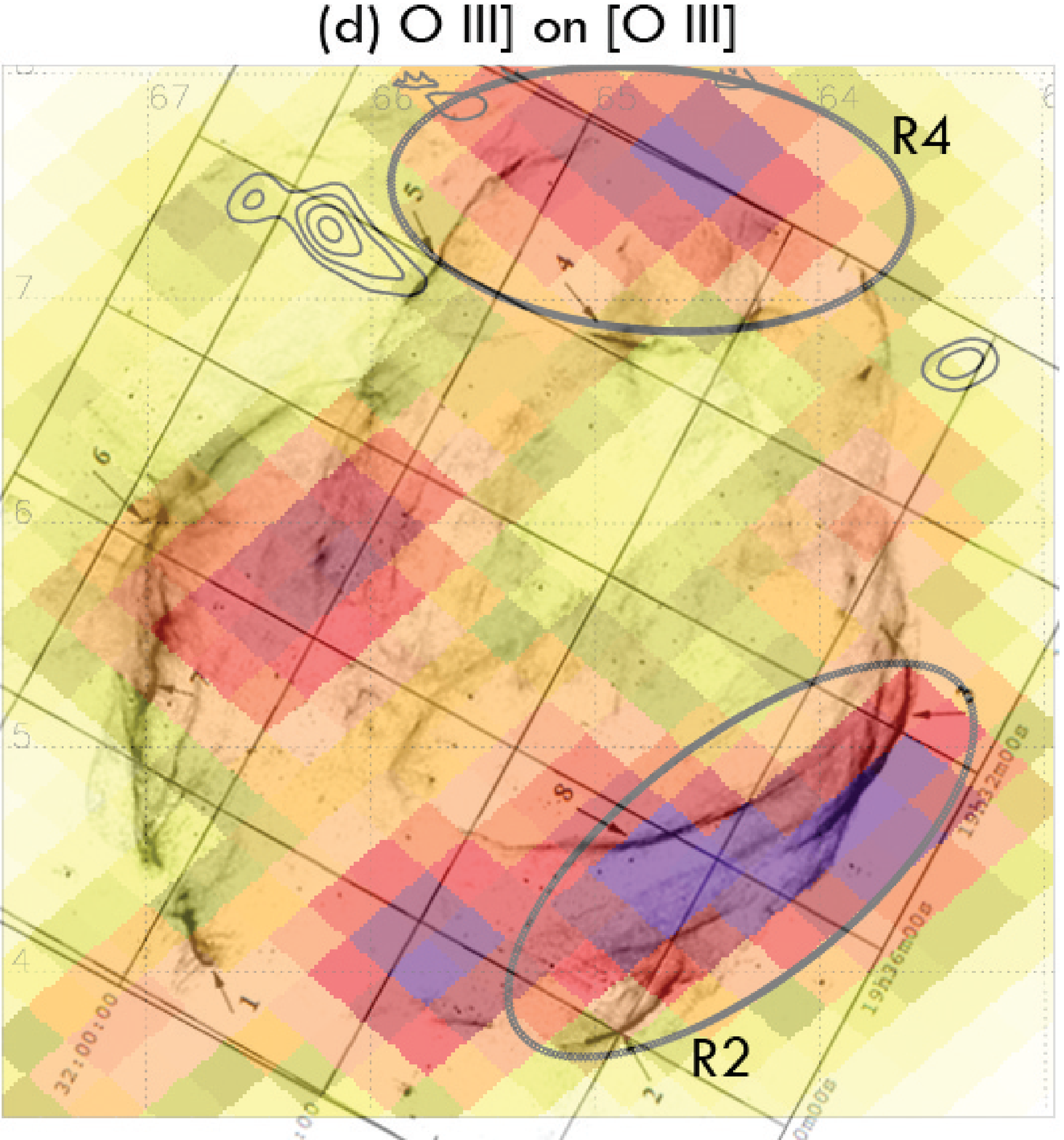}\plotone{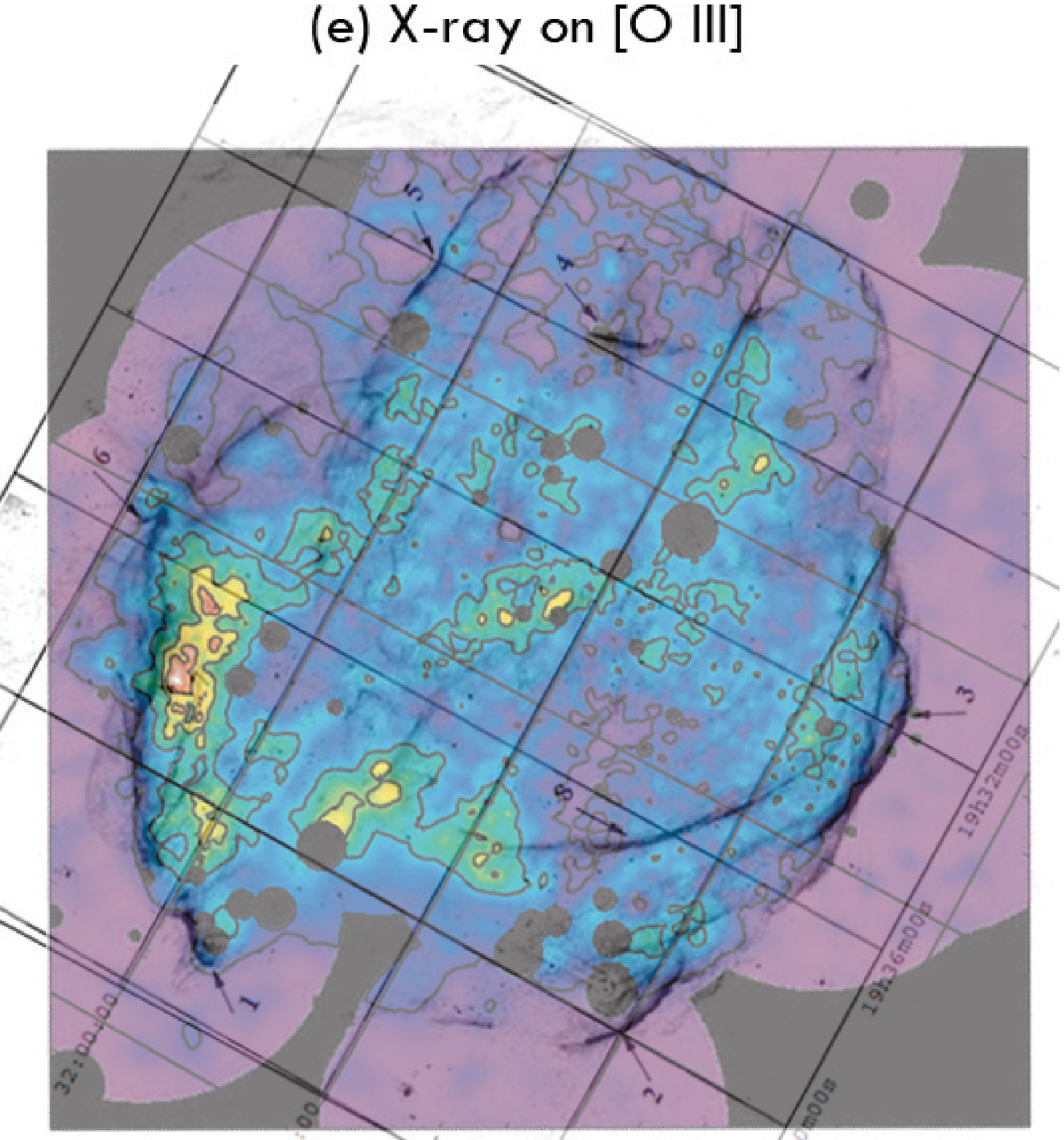}\quad\plotone{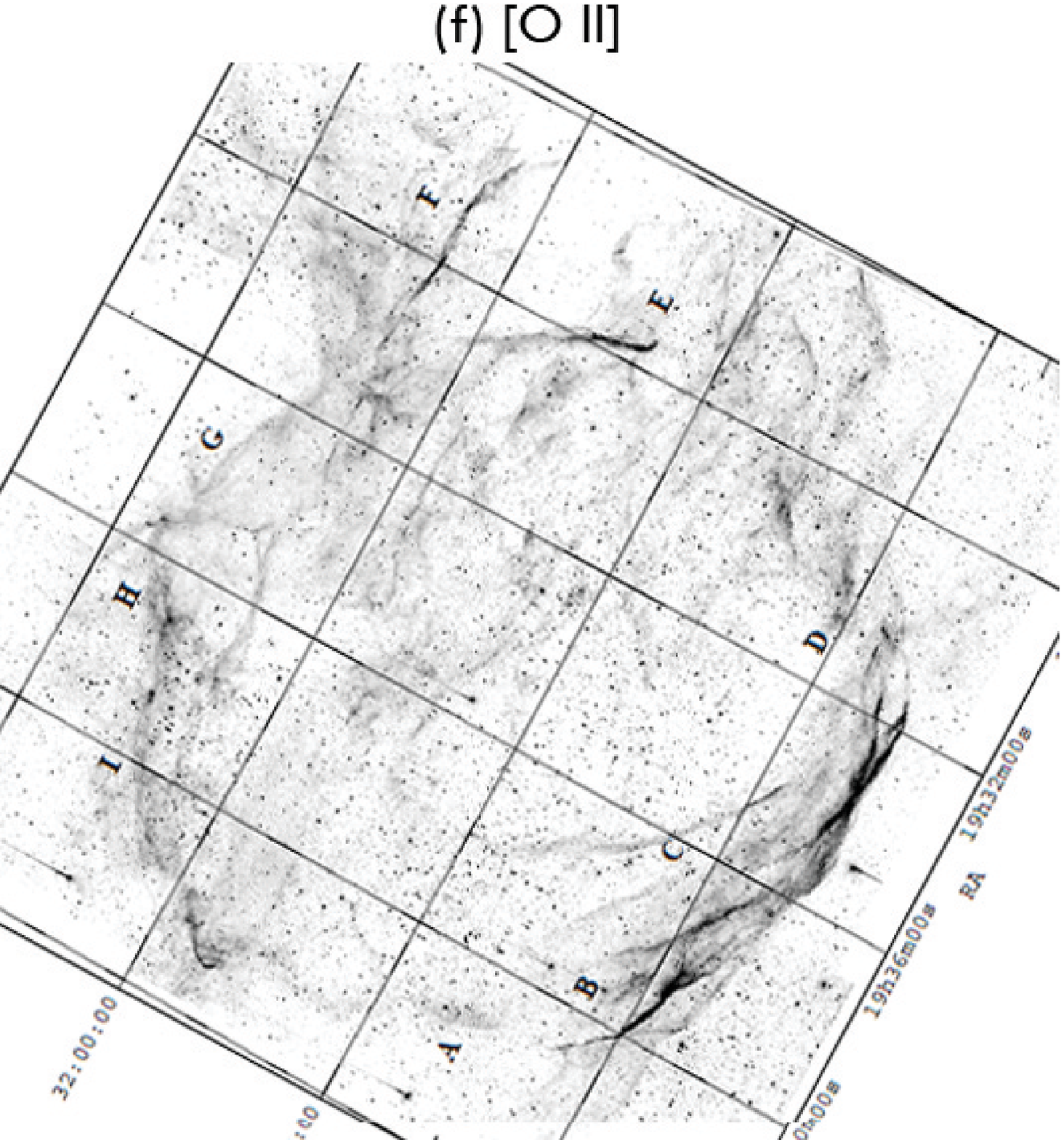} \caption{(a) \ion{C}{4} $\lambda\lambda$1548, 1551, (b) H$_{2}$ $\lambda$1608, (c) \ion{He}{2} $\lambda$1640, (d) \ion{O}{3}] $\lambda\lambda$1661, 1666, and (e) X-ray (0.11--0.284 keV) color-scale images overlaid on [\ion{O}{3}] $\lambda$5007 grey-scale images, as well as (f) [\ion{O}{2}] $\lambda$3727 image. The color-scale in (a)--(d) is the same as Figure 1. The contours in (a)--(d) are dust data, representing $E(\bv)$ = 0.33--0.41 magnitudes with 0.02 intervals. Only the dust contours above $b = 6.5\arcdeg$ are plotted to avoid the complexity of the low-latitude features. The spectra in Figure 3 are extracted from the four subregions designated by R1 through R4 in (a), (b), and (d). The [\ion{O}{3}] $\lambda$5007 and the [\ion{O}{2}] $\lambda$3727 images are from \citet{mav02}. The X-ray images are from \citet{she04}.\label{fig2}}
\end{figure}

\begin{figure}
\epsscale{1} \plotone{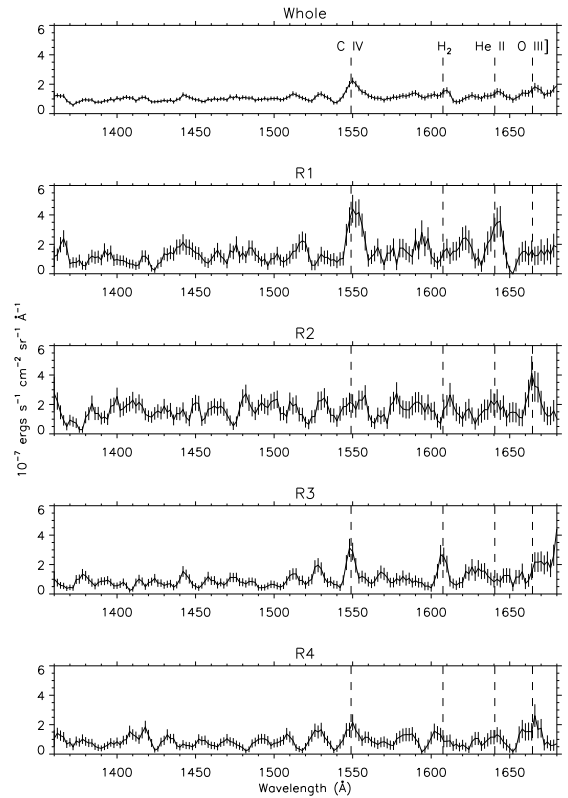} \caption{FIMS/SPEAR L-channel spectra (with 1 $\sigma$ error bars) from the whole region and the four subregions indicated in Figure 2. The spectra are binned at 2 \AA{} intervals and smoothed with a boxcar average of 3 bins. The positions of the identified emission lines are indicated by dashed lines. \label{fig3}}
\end{figure}

In Figures 2(a)--(e), the FUV emission-line images in Figure 1 and the X-ray (0.11--0.284 keV) color-scale image \citep{she04} are overlaid on the [\ion{O}{3}] $\lambda$5007 grey-scale images \citep{mav02}. Additionally, an image of [\ion{O}{2}] $\lambda$3727 \citep{mav02} is shown in Figure 2(f). Dust contours \citep{sch98} are also plotted in (a)--(d) for only the regions of $>$$6.5\arcdeg$ galactic latitude to avoid the complexity of low-latitude features. We marked four interesting regions with designations R1 through R4 in Figures 2(a), (b), and (d): the east \ion{C}{4} region (R1), the southwest \ion{O}{3}] region (R2), the central \ion{C}{4} region including the H$_{2}$-detected regions (R3), and the north \ion{O}{3}] region (R4). In Figure 2(a), the \ion{C}{4} morphology appears in general to correlate with the [\ion{O}{3}] filaments though the \ion{C}{4} emission is quite weak (S/N $<$1.5) on the strong southwest [\ion{O}{3}] filaments and even absent on the northeast [\ion{O}{3}] filament. Interestingly, a dust feature coincides well with no \ion{C}{4} emission region in the northeast. The brightest \ion{C}{4} filamentary feature in the east (R1) is running along the east [\ion{O}{3}] filament and peaks inside it although the boundary of the \ion{C}{4} feature is a little ambiguous due to smoothing. In fact, the \ion{C}{4} peak coincides with the X-ray peak just inside the east [\ion{O}{3}] filament in Figure 2(e). The other bright \ion{C}{4} filamentary feature in the central region coincides well with a fainter [\ion{O}{3}] filament and also has a fainter X-ray counterpart. In Figure 2(b), there are two bright H$_{2}$ regions with S/N $>$1.5. The weaker overlaps with the central \ion{C}{4} and [\ion{O}{3}] filaments, and the stronger in the west has very faint \ion{C}{4} and [\ion{O}{3}] counterparts. In Figure 2(c), the bright \ion{He}{2} region covers the east side of the [\ion{O}{3}] features. Interestingly, the brightest \ion{He}{2} part coincides with the X-ray peak in Figure 2(e). In Figure 2(d), the brightest \ion{O}{3}] filamentary feature in the southwest (R2) overlaps well with the strong [\ion{O}{3}] filaments. This brightest \ion{O}{3}] part also coincides with the brightest [\ion{O}{2}] filaments in Figure 2(f). The other \ion{O}{3}] filamentary feature in the north (R4) seems to extend outside the [\ion{O}{3}] observation region.

\begin{deluxetable*}{cccccccc}
\tablecaption{FUV line luminosities and comparison with the Cygnus Loop and the Vela \label{tbl-1}}
\tablehead{
\colhead{Species} & \colhead{R1} & \colhead{R2} & \colhead{R3} & \colhead{R4} & \colhead{Whole} & \colhead{Cygnus Loop} & \colhead{Vela}
}
\startdata
\ion{C}{4} $\lambda\lambda$1548, 1551 ....& 1.6$\pm$0.6 & $<$0.6 & 1.4$\pm$0.4 & 0.7$\pm$0.3 & 4.5$\pm$1.0 & 4.47$\pm$0.14\tablenotemark{a} & 8.0\tablenotemark{b} \\
H$_{2}$ $\lambda$1608 ...................& $<$0.2 & $<$0.2 & 1.2$\pm$0.4 & $<$0.6 & $<$1.8 & \nodata & \nodata \\
\ion{He}{2} $\lambda$1640 ...............& 1.5$\pm$0.7 & $<$0.8 & $<$0.2 & $<$0.6 & $<$1.4 & 0.68$\pm$0.06\tablenotemark{a} & 1.3\tablenotemark{b} \\
\ion{O}{3}] $\lambda\lambda$1661, 1666 ...& $<$0.8 & 1.4$\pm$0.7 & $<$0.4 & 1.3$\pm$0.5 & $<$3.3 & 0.65$\pm$0.08\tablenotemark{a} & 1.5\tablenotemark{b} \\
X-ray .........................& \nodata & \nodata & \nodata & \nodata & 0.082\tablenotemark{c} & 3.59\tablenotemark{d} & 2.2\tablenotemark{e} \\
\enddata
\tablenotetext{a}{The values are from \citet{seon06}.}
\tablenotetext{b}{The values are from \citet{nis06}.}
\tablenotetext{c}{The value is from \citet{she04} and the corresponding energy band is 0.11--2.04 keV.}
\tablenotetext{d}{The value is from \citet{seon06} and the corresponding energy band is 0.1--4.0 keV.}
\tablenotetext{e}{The value is from \citet{lu00} and the corresponding energy band is 0.1--2.5 keV.}
\tablecomments{R1 through R4 correspond to regions shown in Figure 2. Line luminosities are in units of 10$^{35}$ ergs s$^{-1}$ and calculated adopting a distance of 900 pc \citep{rei79}. Line luminosities with $<$2 $\sigma$ significance are indicated by 1 $\sigma$ upper limits.}
\end{deluxetable*}

In Figure 3, we have plotted the spectra for the whole region and subregions R1 through R4. The spectra were binned at 2 \AA{} intervals and smoothed with a boxcar average of 3 bins. The detector background was subtracted in these spectra despite their small size: $\sim$2\% of the continuum level of the whole region (0.006 counts s$^{-1}$ \AA$^{-1}$; Seon et al. 2010). Most of the FUV continuum is scattered light of the FUV stellar radiation by the interstellar dust \citep{seo10}. The spectrum for the whole region shows the \ion{C}{4} $\lambda\lambda$1548, 1551 doublet, which is unresolved, together with marginal signs of the H$_{2}$ $\lambda$1608, \ion{He}{2} $\lambda$1640, and \ion{O}{3}] $\lambda\lambda$1661, 1666 lines. The emission line features are more clearly seen in the spectra of subregions: \ion{C}{4} and \ion{He}{2} in R1, \ion{O}{3}] in R2, and \ion{C}{4} and H$_{2}$ in R3. For each subregion, we fitted the spectrum to obtain the emission line intensities. For reddening correction, we adopted a foreground neutral hydrogen column density ($N_{\mathrm H}$) of 1.0 $\times$ 10$^{20}$ cm$^{-2}$, the average value obtained from the X-ray spectral fitting in \citet{she04}. Using $E(\bv)$ = 0.017 (from the relation $N_{\mathrm H}$ = 5.8 $\times$ 10$^{21}$ $E(\bv)$ cm$^{-2}$ mag$^{-1}$ of \citet{boh78}) and the extinction curve of \citet{car89} with $R_V$ = 3.1, the reddening-corrected values were obtained. Adopting a distance of 900 pc to the target \citep{rei79}, the line luminosities were calculated and listed along with the values of the \object{Cygnus Loop} \citep{seon06} and the \object{Vela} \citep{nis06} in Table 1. The X-ray luminosities \citep{she04,seon06,lu00} are also shown for comparison. Line luminosities with $<$2 $\sigma$ significance are indicated by 1 $\sigma$ upper limits. As expected from Figure 1, the \ion{C}{4} line is detected at $>$2 $\sigma$ significance in all regions except for R2, and the \ion{O}{3}] line is detected at $>$2 $\sigma$ significance in R2 and R4. The \ion{He}{2} line in R1 and H$_{2}$ $\lambda$1608 line in R3 are also detected at $>$2 $\sigma$ significance.

\section{DISCUSSION}

Because the \ion{C}{4} $\lambda\lambda$1548, 1551 doublet is a resonance line, it is affected by resonant scattering, leading to large attenuation when a shock is viewed edge-on \citep{ray80}. On the other hand, the [\ion{O}{3}] $\lambda$5007 forbidden line is not affected by resonant scattering. Thus, the \ion{C}{4} image overlaid on the [\ion{O}{3}] image in Figure 2(a) can provide clues to the radial structure of this remnant, based on the fact that the \ion{C}{4} line is formed at roughly the same temperature as the [\ion{O}{3}] line \citep{ray97}. In the figure, the \ion{C}{4} emission is much fainter around the southwest [\ion{O}{3}] filaments than around the east and central [\ion{O}{3}] filaments. This suggests that the column density of C$^{+3}$ gas along the line of sight toward the southwest filaments is much larger than toward the other filaments and thus large resonant scattering of the \ion{C}{4} emission occurs in the southwest region. The interpretation is compatible with the previous results obtained from the radio continuum morphologies and the [\ion{O}{3}] line profiles on its filaments. High spatial resolution images of radio continuum at two frequencies (4.8 and 2.6 GHz) shown in Figure 2 of \citet{xia09} appear to be sharpest in the southwest, implying the largest column density in that region. In \citet{bou04}, the [\ion{O}{3}] line profiles for the east and the southwest filaments show approaching (-15, -19 km s$^{-1}$) and receding (0 km s$^{-1}$) radial velocities, respectively, with respect to the given systemic radial velocity (-7 km s$^{-1}$; heliocentric). Hence, the southwest filaments located on the rear side of the remnant likely have a larger path length and column density of emitting gas than the east filaments located on the front side. On the other hand, another [\ion{O}{3}] filament without the \ion{C}{4} emission seen in the northeast region in Figure 2(a) has a diffuse radio continuum counterpart \citep{xia09} and an approaching radial velocity \citep{bou04}. Thus, the weakness or absence of \ion{C}{4} emission in this region is likely due to another cause, that is, heavy interstellar extinction by a foreground dust cloud indicated by the contours in Figures 2(a)--(d).

The FUV emission-line morphologies together with the optical and X-ray morphologies reveal various shocks in different evolutionary stages, as discussed in the following. This indicates that \object{G65.3+5.7} is evolving globally in a transitional stage from the adiabatic to the radiative stage. To see this, we compare the FUV emission lines detected in subregions R1 and R2 with shock models. In the east region (R1), the \ion{C}{4} and \ion{He}{2} emission lines are dominant. The predominance of these two lines is typical of nonradiative shocks in this wavelength range \citep{ray83,lon92,hes94,san00}. The coinciding X-ray filamentary peak seen in Figure 2(e) also supports the existence of a fast shock there. We therefore conclude that the \ion{C}{4} and \ion{He}{2} emission lines detected in subregion R1 are likely from a nonradiative shock right inside the east [\ion{O}{3}] filament which has evolved into an incomplete radiative shock just recently (low degree of completeness). Very high [\ion{O}{3}]/H$\beta$ ratios (18--110) observed along the east [\ion{O}{3}] filament \citep{fes83,mav02} support that the shocks in subregion R1 have a low degree of completeness. This nonradiative shock next to the approaching east [\ion{O}{3}] filament, as mentioned above, is also likely on the front side of the remnant, which makes no significant resonant scattering of the \ion{C}{4} emission. Quantitatively, as indicated in Table 1, the \ion{He}{2} line intensity comparable to the \ion{C}{4} line intensity for subregion R1 is expected when the amount of material swept up by a shock is not large in a nonradiative shock \citep{hes94}. The other \ion{C}{4} filamentary feature with fainter [\ion{O}{3}] and X-ray counterparts around the center of the remnant seems to be an incomplete (probably with a low degree of completeness) radiative shock viewed nearly face-on.

According to an incomplete radiative shock model \citep{ray88}, the [\ion{O}{2}] $\lambda\lambda$3727, 3729/[\ion{O}{3}] $\lambda\lambda$4959, 5007 ratio increases from 0.4 to 1.8 as the [\ion{O}{3}]/H$\beta$ ratio decreases from 110 to 6.4 for a 120 km s$^{-1}$ or 140 km s$^{-1}$ shock. Thus, the brightest [\ion{O}{2}] intensity in the southwest region in Figure 2(f) likely results from the high degree of completeness of shocks in the region, as the [\ion{O}{3}]/H$\beta$ ratios observed along the southwest [\ion{O}{3}] filaments are low (6.4--25) relative to other regions \citep{mav02}. On the other hand, the \ion{O}{3}] $\lambda\lambda$1661, 1666/[\ion{O}{3}] $\lambda\lambda$4959, 5007 ratio is always around 0.3 when the [\ion{O}{3}]/H$\beta$ ratio is in the range of 6.4--110 according to the same model \citep{ray88}. Thus, the enhancement of the \ion{O}{3}] emission line in the southwest region (R2) results not from a high degree of completeness but from the largest column density of emitting gas toward subregion R2, as revealed above from the comparison between the \ion{C}{4} and [\ion{O}{3}] morphologies. The largest column density toward subregion R2 seems to make the \ion{O}{3}] lines enhance to a detectable level in only that region. We therefore conclude that the \ion{O}{3}] emission line detected in subregion R2 is from nearly complete radiative shocks, which have the largest column density of emitting gas along the line of sight. According to the same model \citep{ray88}, the \ion{C}{4}/\ion{O}{3}] and \ion{He}{2}/\ion{O}{3}] ratios are predicted to be about 10--13 and 0.4--0.5, respectively, when the [\ion{O}{3}]/H$\beta$ ratio is 14--17 for a 120--140 km s$^{-1}$ shock. In Table 1, the 1 $\sigma$ upper limit of \ion{He}{2} for subregion R2 is compatible with the model. However, the 1 $\sigma$ upper limit of \ion{C}{4}, which is less than half of the detected \ion{O}{3}] intensity, implies that the \ion{C}{4} resonance-line intensity given by the model should be reduced by a factor of 20 or more due to the resonant scattering. Notably, \object{G65.3+5.7} shows an evolutionary asymmetry between the east and the southwest sides in terms of Galactic coordinates in addition to the morphological axisymmetry with a northwest-southeast axis argued in \citet{bou04}. The east-southwest evolutionary asymmetry is likely due to a global density gradient toward the Galactic plane because the southwest side shows no hints of distorted morphologies, which are expected when there are large inhomogeneities in ambient ISM or a remnant is in direct contact with other objects as in the case of the Monogem ring \citep{kim07}.

Subregion R3 was assigned a bit arbitrarily by binding the two separate bright H$_{2}$ regions to show the marginally-detected H$_{2}$ $\lambda$1608 line more markedly. \citet{ray81} suggested that fluorescent emission lines of H$_{2}$ may be observable in some SNRs. These lines are induced by a strong \ion{H}{1} Ly$\beta$ emission line of SNRs and are detectable in ultraviolet and infrared wavelength domains. The strongest fluorescent line in the model is at 1608 \AA{} and the only H$_{2}$ fluorescent line detected in subregion R3 is at the same wavelength. However, they argued that the fluorescent lines more likely arise from molecules formed in the post-shock region than in the pre-shock region, which is not in accord with the above morphological result that the strongest cooling region of \object{G65.3+5.7} is not subregion R3, but subregion R2, as shown in Figure 2. We also found no related features of molecular clouds in subregion R3 from CO survey data \citep{dam01}. As can be seen in Figure 1, the emission patterns for the bright H$_{2}$ $\lambda$1608 regions appear to be dominated by only a few pixels in the two separate regions and there are three bright stars near these pixels. Therefore, the detected H$_{2}$ $\lambda$1608 emission line may come from sources unrelated to \object{G65.3+5.7}. The north \ion{O}{3}] filamentary feature (R4) seen in Figure 2(d) may not be from a related shock as well. \citet{bou04} presented a supplementary image covering the outside of the [\ion{O}{3}] observation region indicated in Figure 2(d), which shows no strong optical features in subregion R4.

As shown in Table 1, total \ion{C}{4} luminosity over the whole region of \object{G65.3+5.7} is about the same order of magnitude as those of the \object{Cygnus Loop} and the \object{Vela} which are in a similar evolution stage (recently reached radiative stage). The 1 $\sigma$ upper limits of \ion{He}{2} and \ion{O}{3}] obtained for the whole region are also within an order of magnitude. On the other hand, the total X-ray luminosity of \object{G65.3+5.7} is much less (by more than an order of magnitude) than those of the \object{Cygnus Loop} and the \object{Vela}, although the ranges of X-ray energy bands are somewhat different. In a hydrodynamic simulation of evolving SNRs \citep{she98,she99}, the soft X-ray luminosity decreases much faster than the \ion{C}{4} luminosity as a SNR evolves in a transitional stage from the adiabatic to the radiative stage (between 1.0 and 5.0 $\times$ 10$^{5}$ yr according to the simulation). Comparing Table 2 of \citet{she98} with Figure 7 of \citet{she99}, the X-ray luminosity decreases by more than an order of magnitude during the period; on the other hand, the \ion{C}{4} luminosity decreases by only about half during the same period. Thus, the result that the X-ray luminosity of \object{G65.3+5.7} is considerably low relative to the \ion{C}{4} luminosity can be explained by the conclusion that it has almost evolved into the radiative stage in the global sense. \citet{she04} concluded that \object{G65.3+5.7} has lost its bright X-ray shell and has a mixed-morphology as it has evolved beyond the adiabatic into the radiative stage. They argued that this oldest known mixed-morphology SNR supports the evolutionary hypothesis for this puzzling class of remnants (mixed-morphology). The above result of the luminosity comparison supports their explanation that the X-ray luminosity of \object{G65.3+5.7} has significantly dimmed.

\section{CONCLUSIONS}

We have presented the first FUV emission-line morphologies of the whole region of the SNR \object{G65.3+5.7} using the data set of FIMS/SPEAR. Although no definite interpretations about the detected H$_{2}$ $\lambda$1608 line have been concluded, the \ion{C}{4} $\lambda\lambda$1548, 1551, \ion{He}{2} $\lambda$1640, and \ion{O}{3}] $\lambda\lambda$1661, 1666 morphologies were found to have close relations with the optical and/or X-ray images. A comparison of the \ion{C}{4} morphology with the optical [\ion{O}{3}] $\lambda$5007 image revealed large \ion{C}{4} resonant-scattering in the southwest region, implying a large column density of emitting gas along the line of sight toward the region. On the other hand, the absence of \ion{C}{4} in the northeast region is likely due to heavy interstellar extinction by a foreground dust cloud. The FUV emission-line morphologies also revealed various shocks in different evolutionary stages. The \ion{C}{4} and \ion{He}{2} emission lines detected in the east region are likely from a nonradiative shock just inside the east optical [\ion{O}{3}] filament. The \ion{O}{3}] emission line detected in the southwest region is from nearly complete radiative shocks where a large column density of emitting gas toward that region makes \ion{O}{3}] detectable. These results confirm that \object{G65.3+5.7} is evolving globally in a transitional stage from the adiabatic to the radiative stage. Also, it is revealed that \object{G65.3+5.7} has evolved asymmetrically between the east and the southwest sides in the Galactic coordinates, possibly due to the global Galactic density gradient. A comparison of total \ion{C}{4} and X-ray luminosities of \object{G65.3+5.7} with those of the \object{Cygnus Loop} and the \object{Vela} SNRs shows that the X-ray luminosity of \object{G65.3+5.7} is considerably low relative to the \ion{C}{4} luminosity. This implies that \object{G65.3+5.7} has almost evolved into the radiative stage in the global sense and supports the previous proposal that \object{G65.3+5.7} has lost its bright X-ray shell and has a mixed-morphology as it has evolved beyond the adiabatic stage.

\acknowledgments

FIMS/SPEAR is a joint project of the Korea Advanced Institute of Science and Technology, the Korea Astronomy and Space Science Institute, and the University of California at Berkeley, funded by the Korean Ministry of Science and Technology and the National Aeronautics and Space Administration (NASA) Grant NAG5-5355. This work was supported by a National Research Foundation of Korea grant funded by the Korean government (grant no. 313-2008-2-C00377).


\begin{thebibliography}{}
\bibitem[Bohlin et al.(1978)]{boh78} Bohlin, R. C., Savage, B. D., \& Drake, J. F. 1978, \apj, 224, 132
\bibitem[Boumis et al.(2004)]{bou04} Boumis, P., Meaburn, J., L\'opez, J. A., Mavromatakis, F., Redman, M. P., Harman, D. J., \& Goudis, C. D. 2004, \aap, 424, 583
\bibitem[Cardelli et al.(1989)]{car89} Cardelli, J. A., Clayton, G. C., \& Mathis, J. S. 1989, \apj, 345, 245
\bibitem[Dame et al.(2001)]{dam01} Dame, T. M., Hartmann, D., \& Thaddeus, P. 2001, \apj, 547, 792
\bibitem[Edelstein et al.(2006a)]{ede06a} Edelstein, J., et al. 2006a, \apjl, 644, L153
\bibitem[Edelstein et al.(2006b)]{ede06b} --------. 2006b, \apjl, 644, L159
\bibitem[Fesen et al.(1985)]{fes85} Fesen, R. A., Blair, W. P., \& Kirshner, R. P. 1985, \apj, 292, 29
\bibitem[Fesen et al.(1983)]{fes83} Fesen, R. A., Gull, T. R., \& Ketelsen, D. A. 1983, \apjs, 51, 337
\bibitem[G\'orski et al.(2005)]{gor05} G\'orski, K. M., Hivon, E., Channelay, A. J., Wandelt, B. D., Hansen, F. K., Reinecke, M., \& Bartelmann, M. 2005, \apj, 622, 759
\bibitem[Gull et al.(1977)]{gul77} Gull, T. R., Kirshner, R. P., \& Parker, R. A. R. 1977, \apjl, 215, L69
\bibitem[Hester et al.(1994)]{hes94} Hester, J. J., Raymond, J. C., \& Blair, W. P. 1994, \apj, 420, 721
\bibitem[Kim et al.(2007)]{kim07} Kim, I.-J., Min, K.-W., Seon, K.-I., Park, J.-W., Han, W., Park, J.-H., Nam, U.-W., Edelstein, J., Sankrit, R., \& Korpela, E. J. 2007, \apjl, 665, L139
\bibitem[Long et al.(1992)]{lon92} Long, K. S., Blair, W. P., Vancura, O., Bowers, C. W., Davidsen, A. F., \& Raymond, J. C. 1992, \apj, 400, 214
\bibitem[Lu \& Aschenbach (2000)]{lu00} Lu, F. J., \& Aschenbach, B. 2000, \aap, 362, 1083
\bibitem[Mavromatakis et al.(2002)]{mav02} Mavromatakis, F., Boumis, P., Papamastorakis, J., \& Ventura, J. 2002, \aap, 388, 355
\bibitem[Nishikida et al.(2006)]{nis06} Nishikida, K., et al. 2006, \apjl, 644, L171
\bibitem[Raymond et al.(1980)]{ray80} Raymond, J. C., Black, J. H., Dupree, A. K., Hartmann, L., \& Wolff, R. S. 1980, \apj, 238, 881
\bibitem[Raymond et al.(1981)]{ray81} Raymond, J. C., Black, J. H., Dupree, A. K., Hartmann, L., \& Wolff, R. S. 1981, \apj, 246, 100
\bibitem[Raymond et al.(1983)]{ray83} Raymond, J. C., Blair, W. P., Fesen, R. A., \& Gull, T. R. 1983, \apj, 275, 636
\bibitem[Raymond et al.(1997)]{ray97} Raymond, J. C., Blair, W. P., Long, K. S., Vancura, O., Edgar, R. J., Morse, J., Hartigan, P., \& Sanders, W. T. 1997, \apj, 482, 881
\bibitem[Raymond et al.(1988)]{ray88} Raymond, J. C., Hester, J. J., Cox, D., Blair, W. P., Fesen, R. A., \& Gull, T. R. 1988, \apj, 324, 869
\bibitem[Reich et al.(1979)]{rei79} Reich, W., Berkhuijsen, E. M., \& Sofue, Y. 1979, \aap, 72, 270
\bibitem[Sabbadin \& D'Odorico (1976)]{sab76} Sabbadin, F., \& D'Odorico, S. 1976, \aap, 49, 119
\bibitem[Sankrit et al.(2000)]{san00} Sankrit, R., Blair, W. P., Raymond, J. C., \& Long, K. S. 2000, \aj, 120, 1925
\bibitem[Schlegel et al.(1998)]{sch98} Schlegel, D. J., Finkbeiner, D. P., \& Davis, M. 1998, \apj, 500, 525
\bibitem[Seon(2006)]{seo06} Seon, K.-I. 2006, J. Korean Phys. Soc., 48, L331 (arXiv:0703168)
\bibitem[Seon et al.(2010)]{seo10} Seon, K.-I., Edelstein, J., Korpela, E. J., Witt, A., Min, K.-W., Han, W., Shinn, J.-H., Kim, I.-J., \& Park, J.-W. 2010, \apjs, submitted (arXiv:1006.4419)
\bibitem[Seon et al.(2006)]{seon06} Seon, K.-I., Han, W., Nam, U.-W., Park, J.-H., Edelstein, J., Korpela, E. J., Sankrit, R., Min, K.-W., Ryu, K., \& Kim, I.-J. 2006, \apjl, 644, L175
\bibitem[Shelton(1998)]{she98} Shelton, R. L. 1998, \apj, 504, 785
\bibitem[Shelton(1999)]{she99} --------. 1999, \apj, 521, 217
\bibitem[Shelton et al.(2004)]{she04} Shelton, R. L., Kuntz, K. D., \& Petre, R. 2004, \apj, 615, 275
\bibitem[Thompson et al.(1978)]{tho78} Thompson, G. I., Nandy, K., Jamar, C., Monfils, A., Houziaux, L., Carnochan, D. J., \& Wilson, R. 1978, Catalogue of Stellar Ultraviolet Fluxes (London: Sci. Res. Council)
\bibitem[Xiao et al.(2009)]{xia09} Xiao, L., Reich, W., F\"urst, E., \& Han, J. L. 2009, \aap, 503, 827
\end{thebibliography}
\end{document}